\documentclass[aps,secnumarabic,nobalancelastpage,
nofootinbib,onecolumn,12pt]{revtex4}%
\usepackage{amsfonts}
\usepackage{amsmath}
\usepackage{graphics}
\usepackage{graphicx}
\usepackage{longtable}
\usepackage{url}
\usepackage{bm}
\usepackage{amssymb}%
\begin{document}
\title{Differential Capacitance of Electric Double Layers: A Poisson-Bikerman Formula}
\author{Ren-Chuen Chen}
\affiliation{Department of Mathematics, National Kaohsiung Normal University, Kaohsiung
802, Taiwan; rcchen@nknucc.nknu.edu.tw}
\author{Chin-Lung Li}
\affiliation{Institute of Computational and Modeling Science, National Tsing Hua
University, Hsinchu 300, Taiwan}
\author{Jen-Hao Chen}
\affiliation{Institute of Computational and Modeling Science, National Tsing Hua
University, Hsinchu 300, Taiwan}
\author{Bob Eisenberg}
\affiliation{Department of Physiology and Biophysics, Rush University, Chicago IL 60612
USA; beisenbe@rush.edu; Department of Applied Mathematics, Illinois Institute
of Technology, Chicago IL 60616 USA}
\author{Jinn-Liang Liu}
\affiliation{Institute of Computational and Modeling Science, National Tsing Hua
University, Hsinchu 300, Taiwan; jinnliu@mail.nd.nthu.edu.tw}
\maketitle

\textbf{Abstract.} We propose a Poisson-Bikerman (PBik) formula for
calculating the differential capacitance (DC) of electrical double layers
(EDLs) in aqueous electrolytes or ionic liquids. The PBik theory is a
generalization of the classical Poisson-Boltzmann theory to include different
steric energies of different-sized ions and water similar to different
electrical energies for different-charged ions. Water and ions with
interstitial voids in this molecular mean field theory have their physical
volumes as they do in molecular dynamics simulations. The PBik formula derived
from Fermi distributions of ions and water in arbitrary shape and volume
reduces to the Bikerman-Freise formula derived from the lattice model of
equal-sized ions. The DC curves predicted by the Gouy-Chapman formula are
U-shaped (for point-like ions with zero volume and very dilute solutions). The
curves change from U shape to camel shape (Bactrian) and then to bell shape
(for finite size ions) as the volume fraction of ions and water changes from
zero to medium value then to large value. The transition is characterized by
critical and inflection voltages in terms of the particle volume fraction.
These voltages determine steric and electrical energies that describe the
space/charge competition and saturation properties of ions and water packed in
the condensed layer of EDLs under high field conditions. Steric energy is as
important as electrical energy in these conditions. PBik computes symmetric DC
curves from delicately balanced steric interactions of asymmetric-size\ ions
and water like the experimental data of KPF$_{6}$ in aqueous solution. It
computes asymmetric\ curves and captures delicately balanced steric or
electrical interactions of ions having different volumes or charges in ionic liquids.

\section{INTRODUCTION}

Differential capacitance (DC) measures the voltage-dependent capacitance of
electrolyte capacitors \cite{Boc70}. It reflects the screening strength of
electrical double layers (EDLs) that change with the surface potential of
electrodes and the composition, concentration, charge, and size of ions and
solvents that screen the potential in a complex way. The EDL determines most
of the properties of solutions because of the enormous strength of the
electric field compared to concentration fields, except in the most crowded
situations. Studies of EDLs are of fundamental importance in adsorbent,
energy, and membrane technologies
\cite{Att96,Bor97,Con99,Kil07,Kor07,Hua08,Baz09,Baz11,Fen12,Kor14,Pil15,Smi16,Cae17,Yu17,Ina18,Fau19,
Lia19,May19,Mat20} as well as in biological systems
\cite{Vla90,And95,Eis96,Hil01,Fog02,Faw04,Kun10,Val18,Liu20}.

Classical theories of EDLs developed by Gouy and Chapman in the 1910s are
based on the Poisson-Boltzmann (PB) equation, where ions are treated as point
charges without volumes with distributions described by Boltzmann statistics
\cite{Kor07,Baz09,Liu20}. These theories predict values of capacitance that
diverge to infinity as the surface potential tends to infinity \cite{Baz09}.
The divergence is not found experimentally and so it is not surprising that
many authors have modified the Gouy-Chapman theory since Stern \cite{Ste24} in
1924. Ions are not points so it is natural to seek a remedy to divergence by
including ionic volumes \cite{Att96,Kor07,Baz09,Pil15,May19,Liu20}. Almost all
these modified models (including those developed in recent years
\cite{Bor97,Kil07,Kor07,Baz09,Baz11}) start with a lattice model of
equal-sized ions proposed by Grimley and Mott \cite{Gri47} in 1947. This is
unfortunate in our view. The assumption of equal size seems to us to produce
degenerate models, because so few ions have equal volume. Unequal size must be
expected to produce layering, as a glance at the relative size of say Na and
Rb shows, and so would have behavior very different from the equal size case.
It seems obvious that ions of different size must be considered if the theory
is to be used for biological or electrochemical systems, and indeed to have
reasonably general validity, because all the solutions of biological interest
and almost all of electrochemical interest include ions of quite different
size. The unequal-sized model by Bikerman \cite{Bik42} in 1942 seems a much
more reasonable place to start. We refer to recent reviews
\cite{Kor07,Baz09,Pil15,May19,Liu20} of PB and modified PB models for
comprehensive surveys and discussions of this topic.

To our knowledge, formulas are absent for the differential capacitance of
solutions with any number of particle species with arbitrary shapes, volumes,
and interstitial voids. We propose here such a formula derived from the
Poisson-Bikerman (-Fermi) theory that yields (a) different steric energies for
different-sized ions and molecular water and (b) different electrical energies
for different-charged ions \cite{Liu20,Liu17}. In this molecular mean field
theory, water and ions have their volumes as they do in molecular dynamics
simulations. Distributions of these particles (ions and water molecules) are
of Fermi type \cite{Kor07}, i.e., ions saturate at large or even infinite
potentials \cite{Liu20,Liu17}.

The correlations produced by the interactions of steric and electrostatic
potentials are well described in this theory over a range of conditions and
length scales, or fits would not be possible \cite{Liu20}. We point out that
higher resolution models (such as in all atom simulations) do not necessarily
capture correlations as well as, let alone better than a mean field theory.
The higher resolution models must actually be shown to capture correlations
correctly in calibrated simulations; it is not obvious that simulations using
periodic boundary conditions and lacking realistic boundary conditions can
capture the correlations produced by the electric field. The electric field
must be described by a continuum model because Maxwell's version of Ampere's
law must include a time-varying electric field that is \textit{not} associated
with mass, and in fact extends into a vacuum reaching to the stars, if not
infinity. Boundary conditions must be specified saying how the electric and
magnetic fields connect to the outside world, or infinity. Thus the electric
field cannot be described without boundary conditions. It cannot be computed
without boundary conditions because mathematically speaking it does not exist
without boundary conditions.

The Poisson-Bikerman (PBik) formula is analytical and semi-analytical for
equal-sized and unequal-sized particles, respectively. Steric energies of
different-sized ions and water molecules are analyzed with consistent
mathematical models, consistently for both continuum and molecular models.
Steric energies are important in dealing with mass conservation, dehydration,
dielectric, energetic, and selectivity properties. Indeed, steric energies are
the main controllers of function in the crucially important calcium channels,
transporters, and probably in the enzymes that control most biological
systems. Steric energy controls function in the same sense that a gas pedal
controls the speed of a car in these systems
\cite{Liu13,Liu13a,Liu14,Liu14a,Liu15,Liu16,Liu18}. We show here the delicate
effects of steric properties of ions and water on differential capacitance.
Tiny changes in steric properties produce large changes in differential
capacitance, or to say the same thing another way, measurements of
differential capacitance can be used to determine the steric potential in some
detail, as it changes with composition, concentration, and potential.

Bazant and co-workers \cite{Kil07,Baz09} and Kornyshev \cite{Kor07}
revitalized the Bikerman-Freise \cite{Fre52} formula \cite{Baz09} which shows
how DC curves change from U shape (predicted by the Gouy-Chapman formula for
very dilute solutions or point-like ions at any concentration) to camel shape
(Bactrian) and then to bell shape (for realistic finite size ions). They
introduced a mean volume fraction of ions to characterize the steric effects
of uniform ions and use it to derive a critical (diffuse-layer) potential and
explain the transition mechanism of these DC shapes \cite{Kil07}. The mean
volume fraction is the product of the size and the bulk concentration of ions.
We generalize the mean volume fraction to ions and water having nonuniform
volumes and arbitrary shapes, where the volumes are physical.

The critical potential approximately separates monotonic and non-monotonic DC
curves as the voltage varies from small to large \cite{Kil07,Baz09}. The
critical potential can also be a voltage at a hypothetical interface between
condensed and diffuse layers of electrical double layers at large voltage
\cite{Kil07,Baz09}. We use our DC formula to define critical and inflection
voltages, where a DC curve attains its extrema and inflection points (at which
the curvature changes its sign), respectively. These voltages give a precise
description of the transitions between condensed and diffuse layers, on the
one hand, and monotonic and non-monotonic DC curves on the other.

The DC formula shows that the symmetric DC curves of experimental data of
KPF$_{6}$ aqueous solution observed by Valette \cite{Val81} arise from
delicately balanced steric interactions of asymmetric-size ions and water.
Asymmetric curves of ionic liquids result from asymmetric ions in size or
charge, and the transitions of DC curves are characterized by the general mean
volume fraction with its critical and inflection voltages.

\section{THEORY}

The total volume of an aqueous electrolyte system with $K$ species of ions in
a solvent domain $\Omega$ is
\begin{equation}
V=\sum_{i=1}^{K+1}v_{i}N_{i}+V_{K+2}, \tag{1}%
\end{equation}
where $K+1$ and $K+2$ denote water and voids, respectively, $v_{i}$ is the
volume of each species $i$ particle, $N_{i}$ is the total number of species
$i$ particles, and $V_{K+2}$ is the total volume of all the voids
\cite{Liu14}. Dividing this volume equation in bulk conditions by $V$, we get
the bulk volume fraction of voids
\begin{equation}
\Gamma^{b}=1-\sum_{i=1}^{K+1}v_{i}c_{i}^{b}=\frac{V_{K+2}}{V}, \tag{2}%
\end{equation}
where $c_{i}^{b}=\frac{N_{i}}{V}$ are bulk concentrations. If the system is
spatially inhomogeneous with variable electric or steric fields, as in
realistic systems, the constants $c_{i}^{b}$ then change to functions
$c_{i}(\mathbf{r})$ for all $\mathbf{r}$ in $\Omega$ and so does $\Gamma^{b}$
to a function of void fractions
\begin{equation}
\Gamma(\mathbf{r})=1-\sum_{i=1}^{K+1}v_{i}c_{i}(\mathbf{r}). \tag{3}%
\end{equation}
We define the concentrations (distributions) of ions and water in $\Omega$ as
\cite{Liu17}
\begin{equation}
c_{i}(\mathbf{r})=c_{i}^{b}\exp\left(  -\beta_{i}\phi(\mathbf{r})+\frac{v_{i}%
}{\overline{v}}S(\mathbf{r})\right)  ,\quad S(\mathbf{r})=\ln\left(
\frac{\Gamma(\mathbf{r})}{\Gamma^{b}}\right)  , \tag{4}%
\end{equation}
for all $i=1,\cdots,K+1$, where $\phi(\mathbf{r})$ is an electrical potential,
$S(\mathbf{r})$ is called a steric potential \cite{Liu17}, $\beta_{i}%
=q_{i}/k_{B}T$ with $q_{i}$ being the charge on species $i$ ions, $q_{K+1}=0$
for water, $k_{B}$ is the Boltzmann constant, $T$ is an absolute temperature,
and $\overline{v}=\sum_{i=1}^{K+1}v_{i}/(K+1)$ is an average volume.

These distributions are of Fermi type (see Appendix A). The steric potential
$S(\mathbf{r})$ depends on $\phi(\mathbf{r})$ and is an entropic measure of
crowding or emptiness of particles at $\mathbf{r}$. If $\phi(\mathbf{r})=0$
and $c_{i}(\mathbf{r})=c_{i}^{b}$ then $S(\mathbf{r})=0$. The factor
$v_{i}/\overline{v}$ shows that the steric energy $\frac{-v_{i}}{\overline{v}%
}S(\mathbf{r})k_{B}T$ of a type $i$ particle at $\mathbf{r}$ depends not only
on $S(\mathbf{r})$ but also on its volume $v_{i}$ similar to the electrical
energy $\beta_{i}\phi(\mathbf{r})k_{B}T$ depending on both $\phi(\mathbf{r})$
and $q_{i}$ \cite{Liu17}. The charge density $\rho(\mathbf{r})=\sum_{i=1}%
^{K}q_{i}c_{i}(\mathbf{r})$ now includes different-sized ions as first
proposed by Bikerman \cite{Bik42}. We extend his work in the following ways.
We (a) introduce water as a molecule, (b) introduce the factor $v_{i}%
/\overline{v}$, (c) introduce the steric potential, and (d) prove the Fermi
distribution. We thus call the resulting equation
\begin{equation}
-\epsilon\nabla^{2}\phi(x)=\rho(x),\;x\in\Omega\tag{5}%
\end{equation}
a Poisson-Bikerman equation \cite{Liu20} in contrast to the classical
Poisson-Boltzmann equation. Different names are needed because these two
equations yield very different distributions (Appendix A). Here,
$\epsilon=\epsilon_{w}\epsilon_{0}$ with $\epsilon_{w}$ being a dielectric
constant (taken as 78.4 at room temperature) of water and $\epsilon_{0}$ the
vacuum permittivity.

To derive the differential capacitance formulas, we consider the domain
$\Omega$ as a half real line. Eq 5 is then a 1D equation with $\mathbf{r}%
=x\in\lbrack0,\infty)$ and two boundary conditions $-\epsilon\left.
\frac{d\phi}{dx}\right\vert _{x=0}=\sigma$ and $\phi(\infty)=0$, where
$\sigma$ is a surface charge density. The differential capacitance of EDLs is
\begin{equation}
C_{v_{i}}=\epsilon\frac{d}{d\phi_{0}}\left\{  -\left.  \frac{d\phi}%
{dx}\right\vert _{x=0}\right\}  , \tag{6}%
\end{equation}
where $\phi_{0}=\phi(0)$ is the surface potential, the subscript $v_{i}$
indicates that $\phi$ depends on $v_{i}$ in eq 5, and $C_{v_{i}}$ is thus
highly nonlinear in $\phi(x)$, $S(x)$, and $v_{i}$. It is very difficult, if
not impossible, to derive an analytical formula for $C_{v_{i}}$. We can derive
(Appendix B) the analytical DC formula
\begin{equation}
C_{\overline{v}}(\phi_{0})=\frac{\pm\overline{C}\sum_{i=1}^{K}q_{i}c_{i}%
^{b}\exp(-\beta_{i}\phi_{0})}{\exp(-S_{0})\cdot\sqrt{-S_{0}}} \tag{7}%
\end{equation}
only for a special case of equal-sized ions and water, i.e., $v_{i}%
=\overline{v},i=1,\cdots,K+1$, where $\pm=\mp sgn(\sigma)$, $\overline
{C}=\sqrt{\frac{\epsilon\overline{v}}{2k_{B}T}}$, $S_{0}=S(\phi(0))$, and eq 3
simplifies to
\begin{equation}
1=\exp(S)\left[  \Gamma^{b}+\sum_{i=1}^{K+1}v_{i}c_{i}^{b}\exp(-\beta_{i}%
\phi)\right]  . \tag{8}%
\end{equation}

The equal size assumption of particles ($\frac{v_{i}}{\overline{v}}=1$) is the
key to obtaining the formula 7 using eq 8 (Appendix B). The assumption yields
the steric potential $S_{0}$, a critical term in eq 7, that separates from
$\phi_{0}$ in eq 8. For different-sized particles ($\frac{v_{i}}{\overline{v}%
}\neq1$), eq 8 becomes
\begin{equation}
1=\exp(S)\left[  \Gamma^{b}+\sum_{i=1}^{K+1}v_{i}c_{i}^{b}\exp\left(
-\beta_{i}\phi+(\frac{v_{i}}{\overline{v}}-1)S\right)  \right]  . \tag{9}%
\end{equation}

We could not derive a formula from this equation like eq 7 because $S$ cannot
separate from $\phi$ in the exponential term when sizes are unequal. The
steric potential $S$ is a function of $\phi$, namely, $S(\phi)$. Eq 8 yields
an explicit (known) function $S(\phi)=-\ln\left[  \Gamma^{b}+\sum_{i=1}%
^{K+1}v_{i}c_{i}^{b}\exp(-\beta_{i}\phi)\right]  $ whereas eq 9 does not,
i.e., $S(\phi)$ and $\phi$ mingle together in a nonlinear (exponential) way
via the volume term $\frac{v_{i}}{\overline{v}}-1$. It is not surprising that
unequal sizes require a special treatment. A glance at the distribution of
ionic charge that can occur if ions have opposite charges and very different
sizes suggests that the reversal in the sign of the potential should be
possible, because reversal in the \textit{net} charge density is clearly
possible \cite{Gue10}. Such disparate behavior is difficult to capture in an
explicit algebraic formula.

We propose an empirical version of eq 7 as
\begin{equation}
C_{v_{i}}(\phi_{0})=\frac{\pm\overline{C}\sum_{i=1}^{K}q_{i}c_{i}^{b}%
\exp(-\beta_{i}\phi_{0}+\hat{S})}{\exp(-S_{0})\sqrt{-S_{0}}}\text{,} \tag{10}%
\end{equation}
which is not analytical but semi-analytical from eq 7 by adding $\hat{S}%
:=\hat{S}(\phi_{0})$, an unknown (implicit) function of $\phi_{0}$ to be
determined by experimental data (Appendix C). The differential capacitance
$C_{v_{i}}(\phi_{0})$ is thus exponentially sensitive to nonuniform $v_{i}$
because $\hat{S}(\phi_{0})\approx(\frac{v_{i}}{\overline{v}}-1)S(\phi_{0})$.
Eq 10 is a way of describing existing data. It can be used for interpolation
(Appendix C) but not for prediction of the differential capacitance under
other conditions. Prediction requires a formula that we do not know how to
derive, or insight into the physics that we do not yet have. Other perhaps
better empirical expressions may be possible, although they are unknown to us.

It is important to check that the general differential capacitance formula 10
that we use reduces to known forms in appropriate special cases. (i) Formula
10 reduces to eq 7 obviously if $\frac{v_{i}}{\overline{v}}=1$ and $\hat{S}%
=0$. (ii) It reduces to the Bikerman-Freise formula
\cite{Kor07,Baz09,Fre52,Kil07} in Appendix D if $K=2$, $q_{1}=-q_{2}=e$ (the
elementary charge), and $v_{1}=v_{2}\neq0$ (1:1 ionic liquids with equal-sized
ions without water). (iii) It reduces to the Gouy-Chapman formula if
$v_{1}=v_{2}=0$ (point charges) \cite{Kor07}. (iv) It reduces to the Debye
capacitance if $\phi_{0}=0$ in addition to those in (iii) \cite{Kor07}.

Kornyshev introduced a mean volume fraction $\gamma=\overline{N}/N$ in his
Poisson-Fermi theory of ionic liquids based on the lattice-gas model
\cite{Kor07}, where $\overline{N}=N_{1}+N_{2}$ is the total number of cations
$N_{1}$ and anions $N_{2}$ in the bulk of a binary ionic liquid and $N$ is the
total number of uniform lattice sites. We show in Appendix D that the general
distribution model in eq 1 reduces to Kornyshev's uniform lattice model as a
special case. Equivalently, the mean volume fraction $\gamma$ is a special
case of the sum of particle volume fractions $\widetilde{\gamma}=\sum
_{i=1}^{K+1}v_{i}N_{i}/V=\sum_{i=1}^{K+1}v_{i}c_{i}^{b}$, i.e.,
\begin{equation}
\widetilde{\gamma}=\frac{v(N_{1}+N_{2})}{(vN)}=\frac{\overline{N}}{N}%
=2vc^{b}=\gamma\text{.} \tag{11}%
\end{equation}
Eq 11 is in fact Bikerman's bulk ionic volume fraction (derived by Bazant and
co-workers as well \cite{Kil07,Baz09}), where $v=6\overline{v}/\pi$ is the
volume of each site (assuming a primitive cubic lattice), $q_{1}=-q_{2}=e$,
$v_{1}=v_{2}=v\neq0$, and $c^{b}=c_{1}^{b}=c_{2}^{b}$.

In Appendix E, we obtain the inflection points (voltages)%
\begin{equation}
\left\vert \phi_{0}^{\ast\ast}\right\vert =-\alpha V_{T}\ln(\overline{v}c^{b})
\tag{12}%
\end{equation}
of $C_{\overline{v}}(\phi_{0})$, i.e., $C_{\overline{v}}^{\prime\prime}%
(\phi_{0}^{\ast\ast})=0$, where $V_{T}=$ $k_{B}T/e$ is the thermal voltage and
$\alpha$ is a factor of the diffuse-layer voltage \cite{Kil07,Baz09} $\phi
_{0}^{d}=-V_{T}\ln(\overline{v}c^{b})$ in $C_{\overline{v}}^{\prime}(\phi
_{0}^{d})\approx C_{\overline{v}}^{\prime}(\phi_{0}^{\ast})=0$. This factor is
only a ratio of inflection $\phi_{0}^{\ast\ast}$ and diffuse-layer $\phi
_{0}^{d}$ voltages without physical meaning. The diffuse-layer $\phi_{0}^{d}$
(critical $\phi_{0}^{\ast}$) voltage approximately (exactly) separates
monotonic and non-monotonic DC curves as $\phi_{0}$ varies from small to large
\cite{Kil07,Baz09}. It is also a hypothetical voltage at the interface between
condensed and diffuse layers of EDLs at large voltage \cite{Kil07,Baz09}. The
inflection voltage $\phi_{0}^{\ast\ast}$ is where a DC curve changes its
curvature provided that $\overline{v}c^{b}\neq0$ and $\overline{v}c^{b}\neq1$.

Therefore, we further show the following. (v) The sum of volume fractions
$\widetilde{\gamma}=\sum_{i=1}^{K+1}v_{i}c_{i}^{b}$ is more general than the
mean volume fraction $\gamma=2vc^{b}$ in ref 5 and $\gamma=\overline{N}/N$ in
ref 6 (Appendix D). It is more general because it includes ions and water
having different volumes, arbitrary shapes, and any number of solution species.

(vi) From eq 12, if $\overline{v}c^{b}\rightarrow0$ (point-like ions
($\overline{v}\rightarrow0$) or infinite dilute solution ($c^{b}\rightarrow
0$)), then $\lim_{\overline{v}\rightarrow0}\left\vert \phi_{0}^{\ast\ast
}\right\vert =$ $\lim_{c^{b}\rightarrow0}\left\vert \phi_{0}^{\ast\ast
}\right\vert =\infty$, i.e., the inflection voltages $\phi_{0}^{\ast\ast}$
tend to $\pm\infty$ (they do not exist). The DC curve $C_{\overline{v}}%
(\phi_{0})$ is therefore U-shaped for all $\phi_{0}$ with only a unique
minimum at $\phi_{0}^{\ast}=0$ (Appendix E) as predicted by the Gouy-Chapman
theory \cite{Kil07,Kor07}.

(vii) If $0<\overline{v}c^{b}<1$, then $0<\left\vert \phi_{0}^{\ast\ast
}\right\vert <\infty$, i.e., there are four finite inflection voltages
$\pm\phi_{0}^{\ast\ast}\neq0$ and hence two more critical voltages $\pm
\phi_{0}^{\ast}\neq0$. It is mathematically impossible to have $\overline
{v}c^{b}=1$, the volume $\overline{v}$ completely filed by ions or water
molecules without voids (Appendix A). Therefore, the DC curve has three
critical points ($\phi_{0}^{\ast}<0$, $\phi_{0}^{\ast}=0$, $\phi_{0}^{\ast}%
>0$) and four inflection points (two for $\phi_{0}^{\ast\ast}<0$ and two for
$\phi_{0}^{\ast\ast}>0$) and hence is camel-shaped. Furthermore, DC curves
change from U shape in (vi) to camel shape \cite{Kil07,Kor07} as $\overline
{v}c^{b}$ changes from zero to a medium number in the interval $\left(
0,1\right)  $.

(viii) If $\overline{v}c^{b}\rightarrow1$, then $\lim_{\overline{v}%
c^{b}\rightarrow1}\left\vert \phi_{0}^{\ast\ast}\right\vert =0=\phi_{0}^{\ast
}$. Again, the inflection voltages do not exist in this case because the limit
value 0 is a critical voltage which yields a unique maximum of a DC curve
(Appendix E). The curve is thus bell-shaped \cite{Kil07,Kor07}. DC curves
hence change from camel shape in (vii) to bell shape as $\overline{v}c^{b}$
changes from the medium number to a larger number closer to one.

\section{RESULTS}

We first report PBik results by formula 10 for different sizes of ions and
water to fit the experimental data of the differential capacitance $C_{v_{i}%
}(\phi_{0})$ from Valette \cite{Val81} for an electrochemical interface
between a single-crystal Ag electrode (crystal plane (110)) and an aqueous
KPF$_{6}$ electrolyte solution with negligible surface adsorption of ions.
These results are particularly significant because they report the properties
of an atomically clean surface in a defined plane of a perfect crystal
structure. Measurements have been plagued for more than a century by surface
contamination (the charge on the surface rapidly attracts charged `dirt' in
the solution). Different planes in a crystal obviously have different
structures and charge distributions (in most cases) \cite{Val81}. Thus they
have different electrical properties and different EDLs. The plane must then
be defined experimentally if results are to be reproducible, and that plane
must be used in the corresponding theory.

The diameters of cation, anion, and water we use are $d_{+}=2.66$,
$d_{-}=4.69$, and $d_{w}=2.8$ {\AA } \cite{Val81,Roh92}, respectively. The
potential shift at the interface is -0.97 V \cite{Val81,Ina18} so we define
$\overline{\phi}_{0}=\phi_{0}-0.97$. Since the dielectric constant of various
ionic liquids is about 15.5 on average \cite{Ryb18}, we use the permittivity
function
\begin{equation}
\epsilon(\overline{\phi}_{0})=\left[  (78.4-15.5)\exp(-\frac{(\overline{\phi
}_{0}+0.97)^{2}}{2\cdot0.1^{2}})+15.5\right]  \epsilon_{0} \tag{13}%
\end{equation}
as a normal distribution of $\overline{\phi}_{0}$ with the mean value -0.97 V
and the standard deviation 0.1 V at the experimental temperature 25 $^{\circ}$C.

Figure 1 presents the PBik curves $C_{v_{i}}(\phi_{0})$ from eq 10 using the
spline method described in Appendix C fit to the experimental data (symbols)
of aqueous KPF$_{6}|$Ag double layers. The bulk concentrations are 2.5, 10,
40, and 100 mM. These curves are nearly symmetrical about $\overline{\phi}%
_{0}=0.97$ ($\phi_{0}=0$) with respect to the surface potential $\overline
{\phi}_{0}$ while the diameters of K$^{+}$ ($v_{+}/\overline{v}=0.39$ in eq 4)
and PF$_{6}^{-}$ ($v_{-}/\overline{v}=2.15$) are very different. The steric
interactions among K$^{+}$, PF$_{6}^{-}$, and water ($v_{w}/\overline{v}%
=0.46$) play a significant role in the inner-layer capacity of ions in EDL
\cite{Val81}. The asymmetry (i.e., inequality) in size of K$^{+}$, PF$_{6}%
^{-}$, and water produces the asymmetry of the \emph{empirical} steric
potential $\hat{S}$ (shown in Figure 2). On the other hand, the
\emph{analytical} steric potential $S_{0}$ profiles are nearly symmetrical
(shown in Figure 3). The DC curves in $\overline{\phi}_{0}$ and also in
$S_{0}$ are nearly symmetrical.

The steric energy at $\overline{\phi}_{0}=-1.5$ V ($-\hat{S}_{\text{K}^{+}%
}k_{B}T$) in Figure 2 is larger than that at $-0.5$ V ($-\hat{S}%
_{\text{PF}_{6}^{-}}k_{B}T$), for example, because it is more crowded near the
electrode for smaller K$^{+}$ ions at $-1.5$ V than for larger PF$_{6}^{-}$
ions at $-0.5$ V. The effective width of the `EDL capacitor' is larger for
larger ions \cite{Kil07,Baz09}. Therefore, the change of $-\hat{S}%
_{\text{PF}_{6}^{-}}k_{B}T$ (the right wing in Figure 2) is larger than that
of $-\hat{S}_{\text{K}^{+}}k_{B}T$ (the left wing) because of more room for
rearranging of ions and water when the capacitance decreases from a global
maximum in Figure 1 to lower values as the voltage varies (from small to
large). $\hat{S}$ also varies with the bulk concentration $c^{b}$ of KPF$_{6}$
as shown in Figure 2.

\begin{figure}[t]
\centering\includegraphics[scale=0.8]{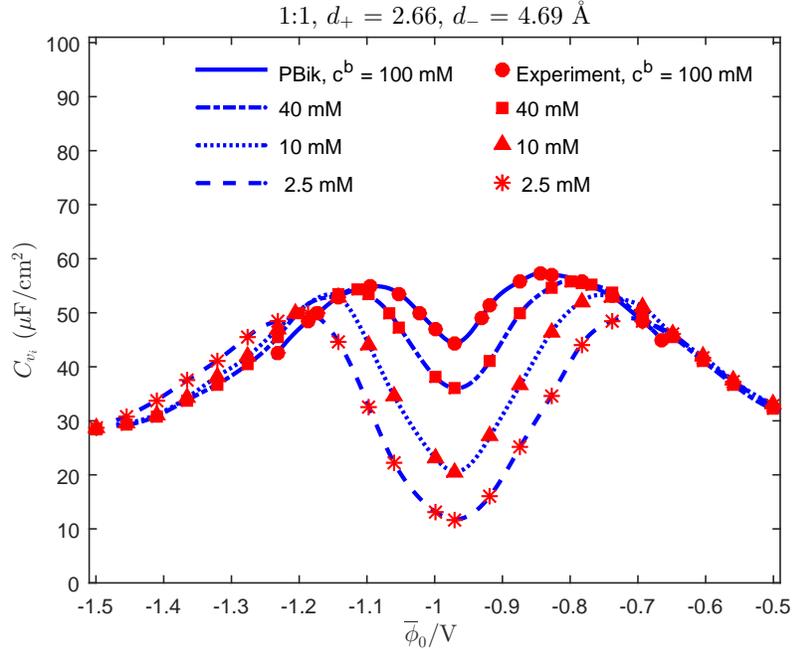}\caption{PBik fitting curves by
eq 10 to the DC experimental data \cite{Val81} (symbols) of aqueous KPF$_{6}%
|$Ag double layers at various concentrations.}%
\end{figure}\begin{figure}[tt]
\centering\includegraphics[scale=0.8]{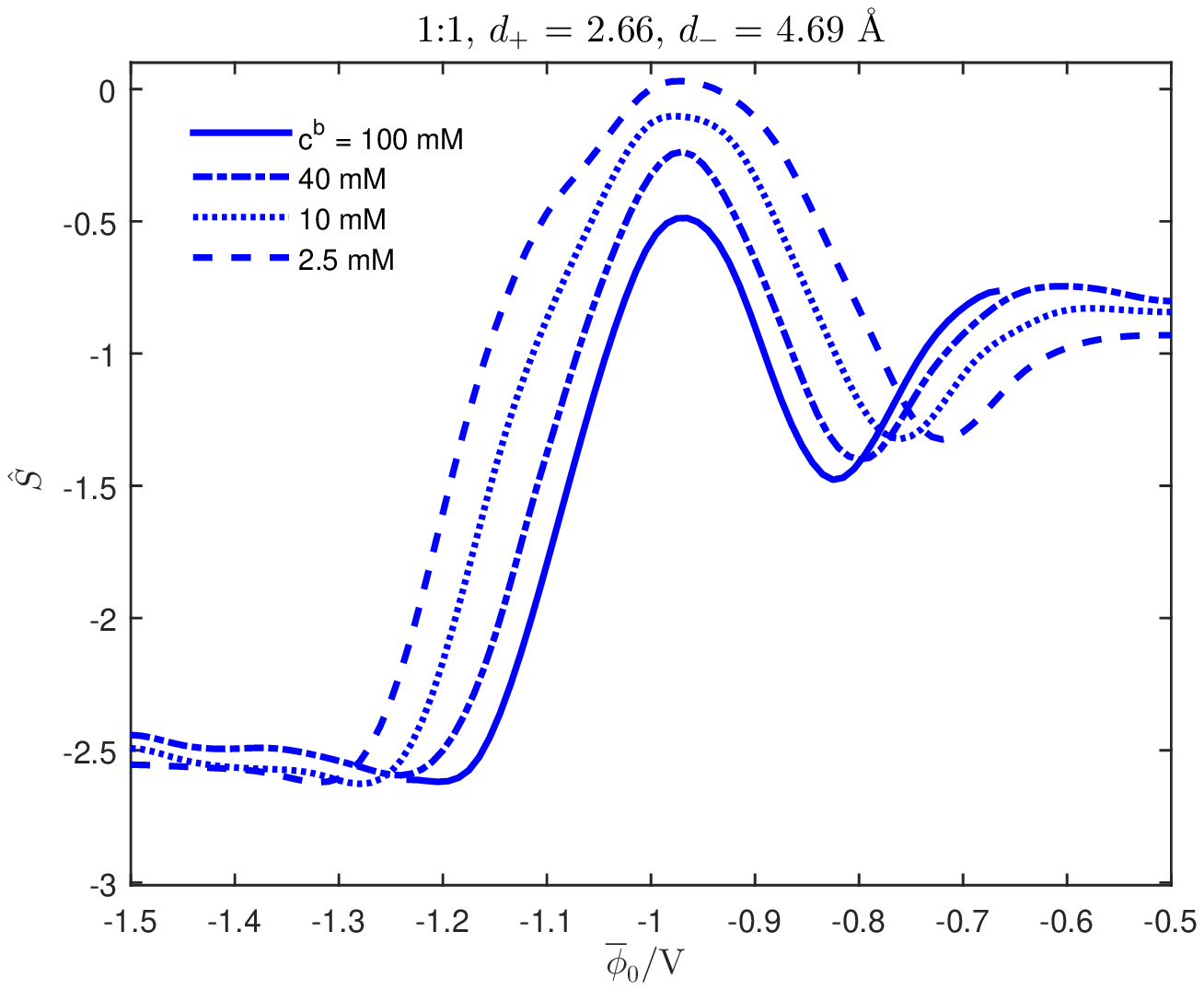}\caption{Asymmetry of the
steric energies $-\hat{S}k_{B}T\approx-\frac{v_{i}}{\overline{v}}S_{0}k_{B}T$
of K$^{+}$ and PF$_{6}^{-}$ in size ($v_{i}$) with respective to
$\overline{\phi}_{0}$ about $\overline{\phi}_{0}=-0.97$ V.}%
\end{figure}\begin{figure}[ttt]
\centering\includegraphics[scale=0.8]{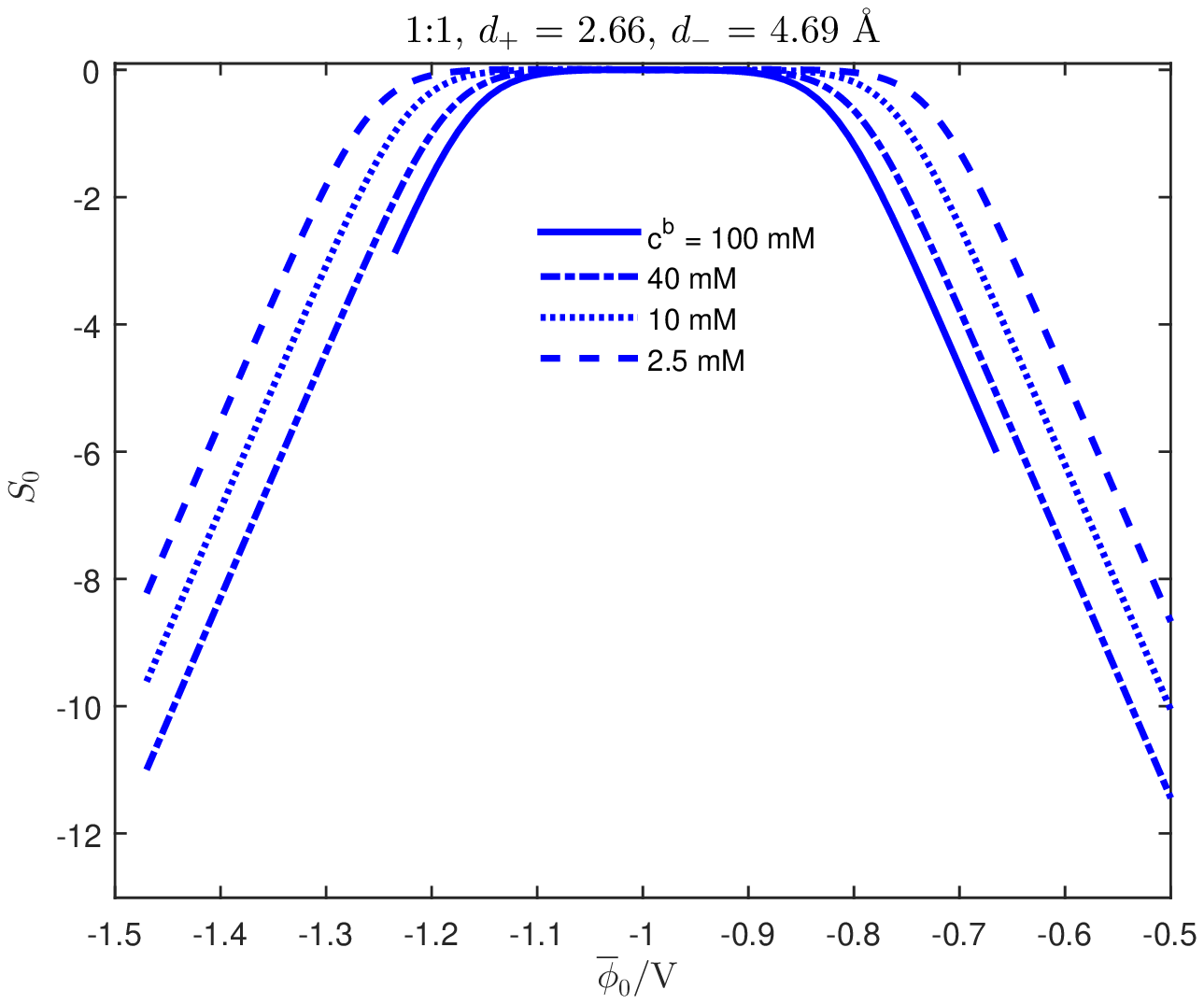}\caption{Symmetry of the steric
potential $S_{0}$ of aqueous KPF$_{6}|$Ag double layers.}%
\end{figure}

Figure 4 shows that the symmetry of Figure 1 is broken when we slightly change
the diameters from $d_{+}=2.66$ to $d_{+}=3$ and $d_{-}=4.69$ to $d_{-}=4$
{\AA }. The same $\hat{S}(\phi_{0})$ in formula 10 is used in both cases. The
results show that $\hat{S}(\phi_{0})$ is very sensitive to ionic diameter due
to the factor $\frac{v_{i}}{\overline{v}}-1$ in eq 9. In this work, $\hat
{S}(\phi_{0})$ is empirical, strongly nonlinear in $\phi_{0}$, and highly
sensitive to the sizes of ions and water. It also depends on the bulk
concentrations of ions and water. It will be interesting to develop analytical
$\hat{S}(\phi_{0})$ in future studies.

\begin{figure}[t]
\centering\includegraphics[scale=0.8]{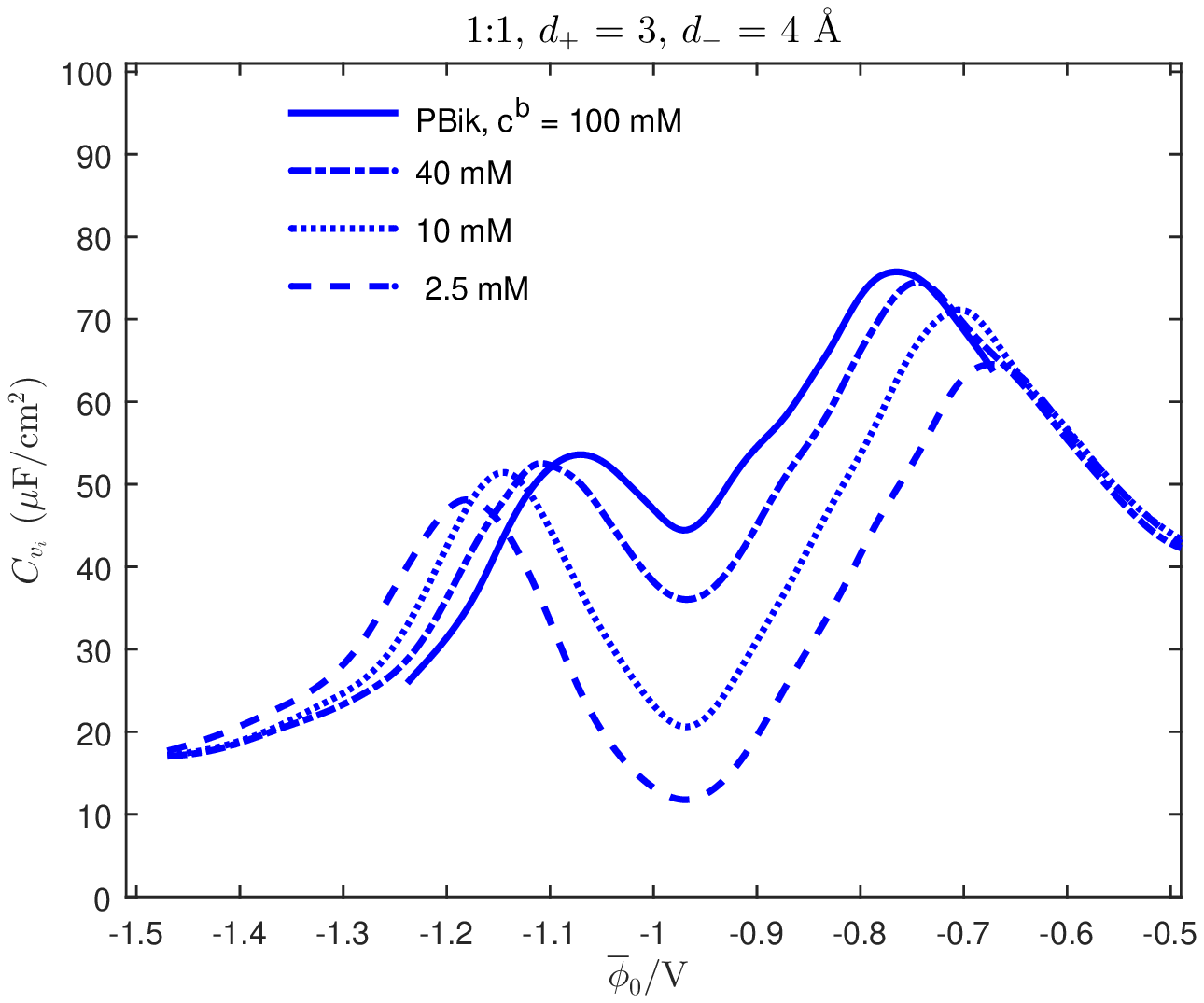}\caption{Broken symmetry of
Figure 1 by slight changes of the diameters $d_{\pm}$.}%
\end{figure}

We next present results using eq 7 for ionic liquids ($K=2$) of equal-sized
ions ($v_{i}=\overline{v}$) with the dielectric constant $\epsilon_{ion}=10$
at $T=$ 25 $^{\circ}$C. The results demonstrate that the analytical formula 7
can successfully describe several important properties of differential
capacitance in electrical double layers. The success is possible because of
the treatment of steric effects of molecules (including molecular water)
within the mean-field framework.

Figure 5 shows two sets of DC curves $C_{\overline{v}}(\phi_{0})$ for 1:1
ionic liquids with (a) $c^{b}=$ 0.01, 0.05, 0.1, 0.5, 1, 2 M at fixed $d_{\pm
}=8$ {\AA } and (b) $d_{\pm}=$ 2, 4, 6, 8, 14, 20 {\AA } at fixed $c^{b}=$
0.02 M, where the capacitance $C_{\overline{v}}$ is scaled by the Debye
capacitance $C_{D}$ and the surface potential $\phi_{0}$ scaled by $V_{T}$.
Figures 5a and 5b clearly demonstrate the transitions of DC curves
$C_{\overline{v}}(\phi_{0})$ from bell shape (with large $c^{b}$ and $d_{\pm}%
$, respectively) to camel shape (medium $c^{b}$ and $d_{\pm}$) then to U shape
(small $c^{b}$ and $d_{\pm}$) as predicted by eq 12.

\begin{figure}[ptbh]
\centering\begin{minipage}[t]{0.49\textwidth}
\centering
\includegraphics[width=8cm]{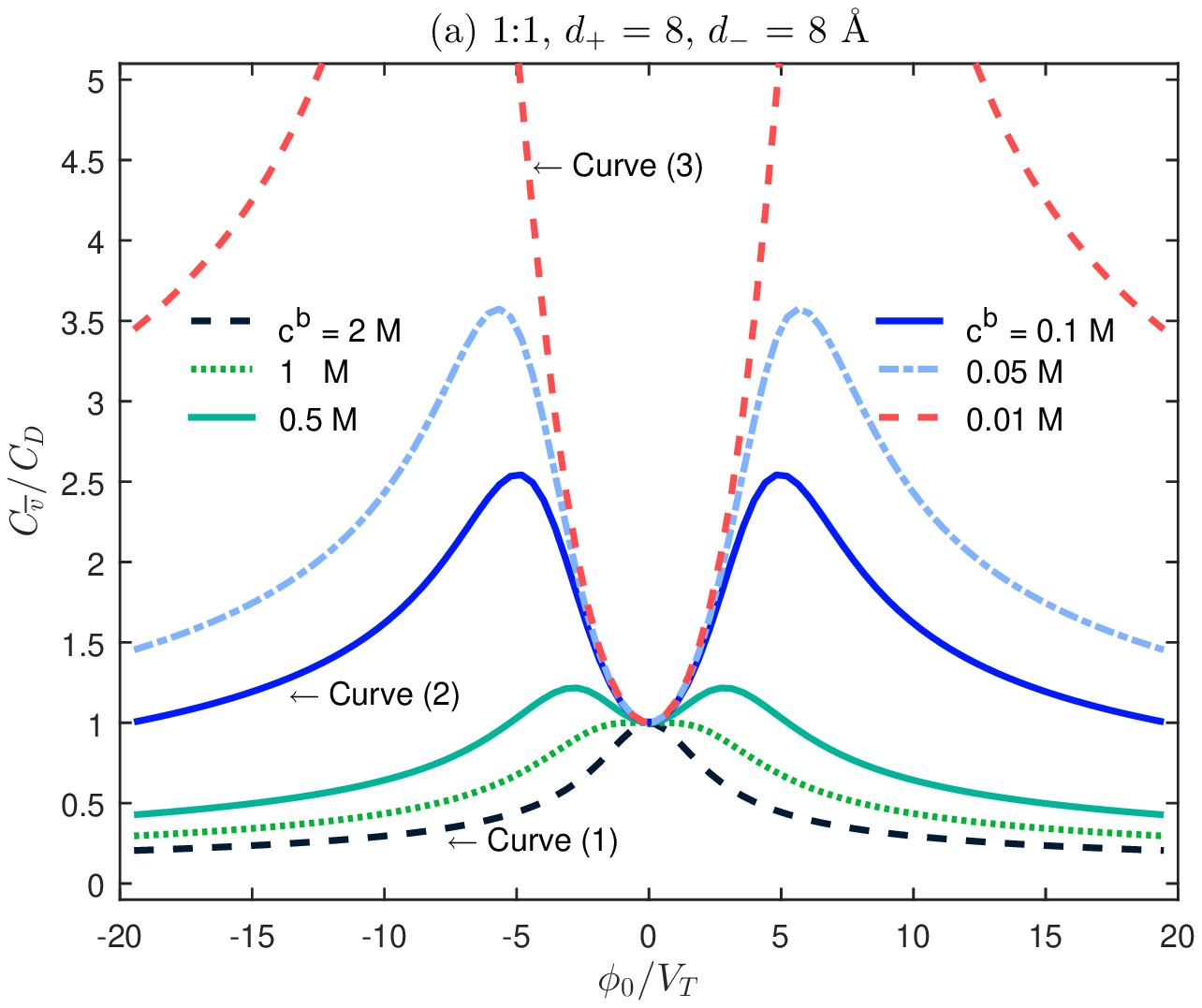}
\end{minipage}\begin{minipage}[t]{0.49\textwidth}
\centering
\includegraphics[width=8cm]{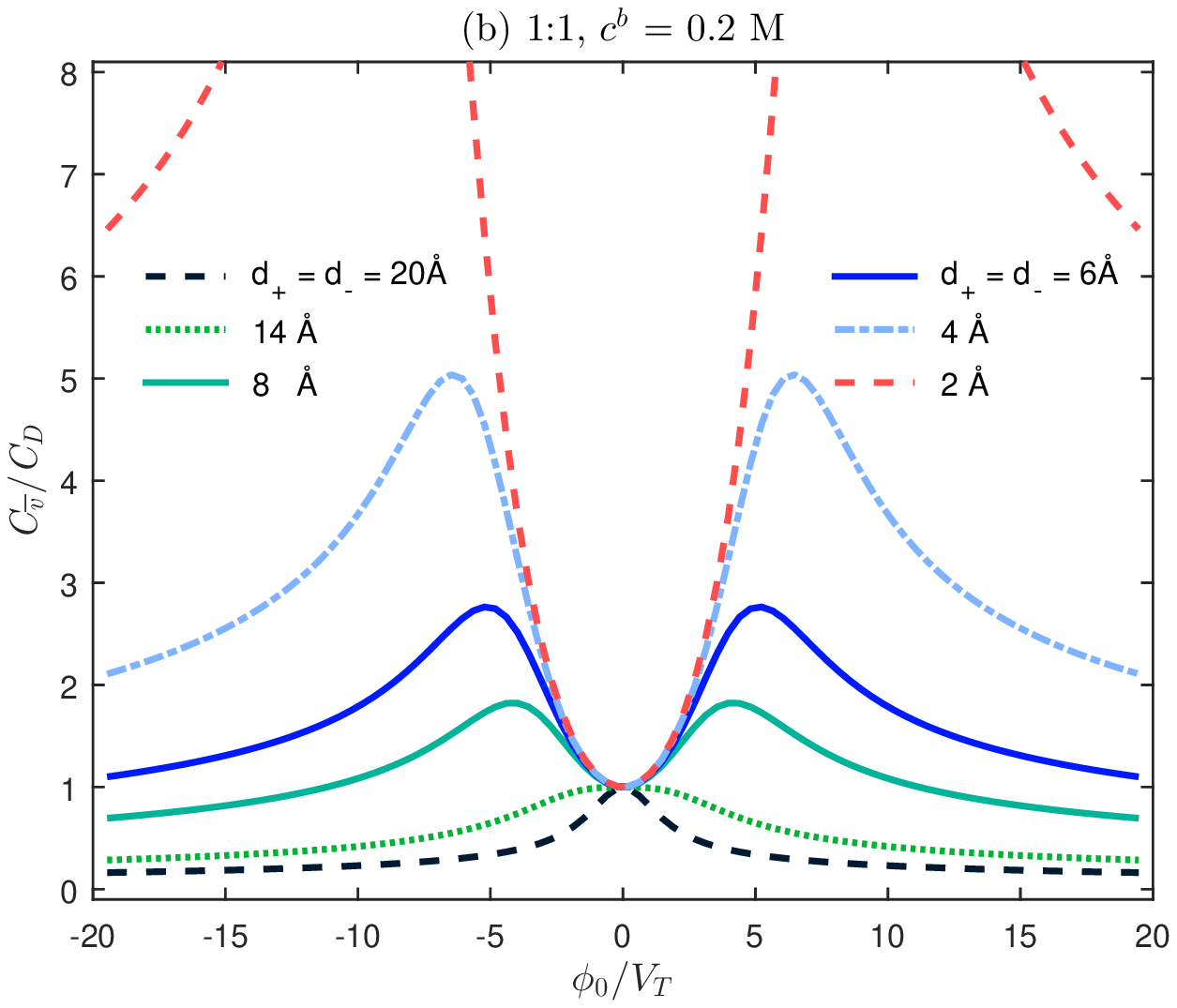}
\end{minipage}
\caption{Transition of DC curves $C_{\overline{v}}(\phi_{0})$ for 1:1 ionic
liquids by eq 12 from bell shape (Curve 1, for example) with large (a) $c^{b}$
and (b) $d_{\pm}$ to camel shape (Curve 2) with medium $c^{b}$ and $d_{\pm}$
then to U (Curve 3) shape with small $c^{b}$ and $d_{\pm}$.}%
\end{figure}

Figure 6 illustrates the main properties of our theory by showing three curves
of the first derivative $C_{\overline{v}}^{\prime}(\phi_{0})$ (Appendix E) for
a 1:1 ionic liquid with $\overline{v}c^{b}=$ (a) $0.323$, (b) $0.0161$, and
(c) $0.0016$. Curves a, b, and c correspond to Curves 1, 2, and 3 in Figure
5a. Curve 1 is bell-shaped because $\overline{v}c^{b}=0.323$ is large for
large ions $\overline{v}$ or large concentration $c^{b}$ as shown in Point
(viii) above. Curves 2 and 3 are camel and U-shaped because $\overline{v}%
c^{b}=0.0161$ and $\overline{v}c^{b}=0.0016$ are medium and small as shown in
Points (vii) and (vi), respectively.

Curves a, b, and c show three sets of reflection voltages ($\phi_{0}^{\ast
\ast}$) \{1.38\}, \{2.87, 6.82\}, and \{5.14, 9.19\} (marked by circles),
which yield \{1.22\}, \{0.69, 1.65\}, and \{0.8, 1.43\} for $\alpha$ in eq 12,
respectively. The critical voltages ($\phi_{0}^{\ast}$) are 0, 4.95, and 7.34
(squares). Voltages are all in $V_{T}$. For Curve b (Curve 2 in Figure 5a),
for example, $\phi_{0}^{\ast}=0$, 4.95; $\phi_{0}^{\ast\ast}=2.87$, 6.82; and
$\alpha=0.69$, 1.65. The ionic liquid with $\overline{v}c^{b}=0.0161$ attains
its maximal capacitance at $\phi_{0}^{\ast}=4.95\approx\phi_{0}^{d}%
=-\ln(\overline{v}c^{b})=4.13$.

\begin{figure}[t]
\centering\includegraphics[scale=0.8]{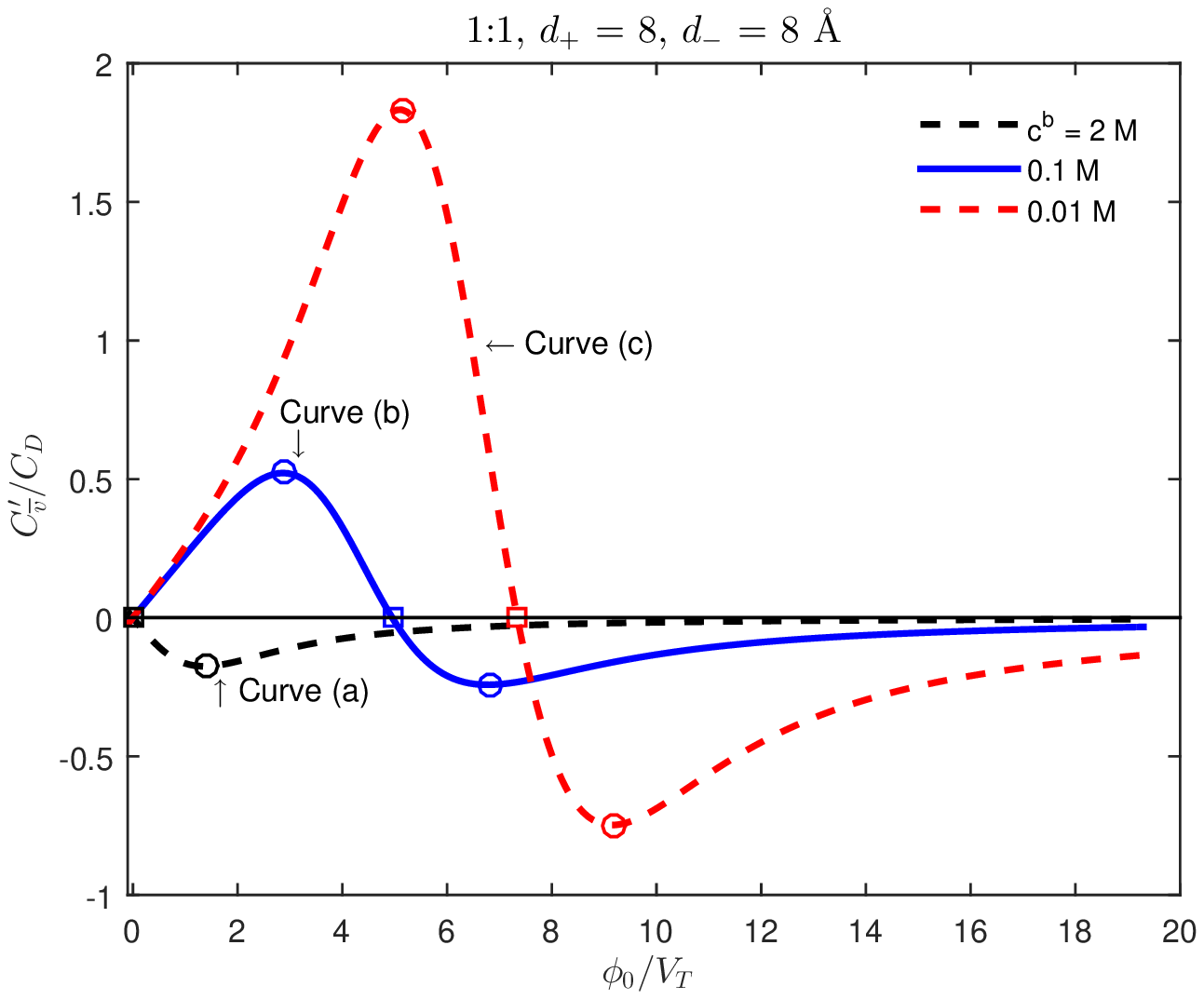}\caption{Critical (squares) and
inflection (circles) voltages determined approximately, i.e., $C_{\overline
{v}}^{\prime}(\phi_{0}^{\ast})=0 $ and $C_{\overline{v}}^{\prime\prime}%
(\phi_{0}^{\ast\ast})=0$, respectively, by the curves of $C_{\overline{v}%
}^{\prime}(\phi_{0})/C_{D}$ with $\overline{v}c^{b}=$ (a) 0.323 (Curve 1 in
Figure 5a), (b) 0.0161 (Curve 2), and (c) 0.0016 (Curve 3).}%
\end{figure}

The critical voltage $\phi_{0}^{\ast}=4.95$ determines a critical steric
potential $S_{0}^{\ast}$ of the ionic liquid that yields the maximal
capacitance
\begin{equation}
C_{\overline{v}}^{\max}(\phi_{0}^{\ast})=\frac{ec^{b}\overline{C}\left[
\exp(\phi_{0}^{\ast}/V_{T})-\exp(-\phi_{0}^{\ast}/V_{T})\right]  }{\exp
(-S_{0}^{\ast})\sqrt{-S_{0}^{\ast}}}=2.5C_{D} \tag{14}%
\end{equation}
shown in Figure 5a and produces a condensed layer \cite{Kil07,Baz09} of ions
packed along the electrode due to the excluded volume $\overline{v}$. When
$\phi_{0}$ increases from $\phi_{0}^{\ast}$ to larger values ($\phi_{0}%
>4.95$), the capacitance $C_{\overline{v}}(\phi_{0})$ decreases from
$C_{\overline{v}}^{\max}(\phi_{0}^{\ast})$, i.e.,%
\[
\frac{C_{\overline{v}}(\phi_{0})}{C_{\overline{v}}^{\max}(\phi_{0}^{\ast}%
)}=\frac{\left[  \exp(\phi_{0}/V_{T})-\exp(-\phi_{0}/V_{T})\right]
\exp(-S_{0}^{\ast})\sqrt{-S_{0}^{\ast}}}{\left[  \exp(\phi_{0}^{\ast}%
/V_{T})-\exp(-\phi_{0}^{\ast}/V_{T})\right]  \exp(-S_{0})\sqrt{-S_{0}}}<1,
\]%
\begin{equation}
\frac{\exp(\phi_{0}/V_{T})-\exp(-\phi_{0}/V_{T})}{\exp(\phi_{0}^{\ast}%
/V_{T})-\exp(-\phi_{0}^{\ast}/V_{T})}<\frac{\exp(-S_{0})\sqrt{-S_{0}}}%
{\exp(-S_{0}^{\ast})\sqrt{-S_{0}^{\ast}}}. \tag{15}%
\end{equation}
Both sides of the inequality 15 are positive and increasing with $\phi_{0}$.
The inequality hence shows that the steric potential $S_{0}$ (crowding
energies) becomes more dominant than the electrical potential $\phi_{0}$
(charging energies) when the surface potential $\phi_{0}$ is greater than the
critical voltage $\phi_{0}^{\ast}$. This implies that the concentration
\begin{equation}
c_{-}(\phi_{0})=c_{-}^{b}\exp\left(  -\beta_{-}\phi_{0}+S_{0}\right)  \tag{16}%
\end{equation}
of packed ions (anions) decreases with $\phi_{0}$ because $S_{0}$ is more
negative yielding larger steric (positive) energies $-S_{0}k_{B}T$ for more
anions to leave the packed region, i.e., the capacitance $C_{\overline{v}%
}(\phi_{0})$ decreases with $\phi_{0}$ as shown in Figure 5a.

The inflection voltage $\phi_{0}^{\ast\ast}=6.82$ describes the saturation
property of the ionic liquid at even greater $\phi_{0}$ because ions have
physical volumes. The steric potential $S_{0}=S(\phi_{0})$ is non-positive and
bounded, i.e., there exists $S_{0}^{\min}=\ln\frac{\Gamma_{0}^{\min}}%
{\Gamma^{b}}$ such that $S_{0}^{\min}\leq S_{0}\leq0$, $\Gamma_{0}^{\min
}=1-\overline{v}c_{-}^{\max}$, and $c_{-}^{\max}<\frac{1}{\overline{v}}$
(Appendix A). From eq 16, we have
\begin{equation}
\lim_{\phi_{0}\rightarrow\phi_{0}^{\max}}c_{-}(\phi_{0})=c_{-}^{b}\exp\left(
-\beta_{-}\phi_{0}^{\max}+S_{0}^{\min}\right)  =c_{-}^{\max}, \tag{17}%
\end{equation}
a saturating concentration with the limits $\phi_{0}^{\max}$ and $S_{0}^{\min
}$ to which the electrical and steric potentials of the system cannot surpass.
This implies that the DC curve $C_{\overline{v}}(\phi_{0})$ approaches a flat
line shown in Figure 5a as $\phi_{0}\rightarrow\phi_{0}^{\max}$. The curve
changes its curvature at $\phi_{0}^{\ast\ast}=6.82$ from concave down in
$2.87<\phi_{0}<6.82$ to concave up in $6.82<\phi_{0}\leq\phi_{0}^{\max}$ so
anions can saturate at $c_{-}^{\max}$ having balanced electrical ($\beta
_{-}\phi_{0}^{\max}k_{B}T$) and steric ($-S_{0}^{\min}k_{B}T$) energies. The
electrical and steric energies at $\phi_{0}=20V_{T}$ in Figure 5a are
$-20k_{B}T$ and $15.87k_{B}T$, respectively. Steric energy is as important as
electrical energy in this condition. These two energies are in fact competing
with each other in the high field (packed) region similar to the space/charge
competition in the binding site of biological ion channels crowded with ions
\cite{Liu20,Non98,Eis03,Eis11}.

Finally, Figure 7 shows that DC curves are asymmetric for ions with unequal
charges in (a) 2:1, 1:2 and (b) 4:1, 1:4 ionic liquids at various
concentrations and fixed $d_{\pm}=4$ {\AA }. The more the charge difference is
between ions (4:1 vs 2:1 for example), the larger the difference is between
global maximal and minimal capacitance (800 vs 400 for $C_{\overline{v}}$ in
Figures 7b and 7a). The maximal capacitance doubles (800 vs 400) if the charge
of ions doubles (4:1 vs 2:1). Comparing green and red curves in Figure 7a
shows the underlying qualitative principle. The more the concentration of an
ionic liquid is ($c_{+}^{b}=1$ vs $c_{+}^{b}=0.001$), the smaller the
capacitance difference is and the narrower between two peaks of a curve (green
vs red).

\begin{figure}[t]
\centering\begin{minipage}[t]{0.49\textwidth}
\centering
\includegraphics[width=8cm]{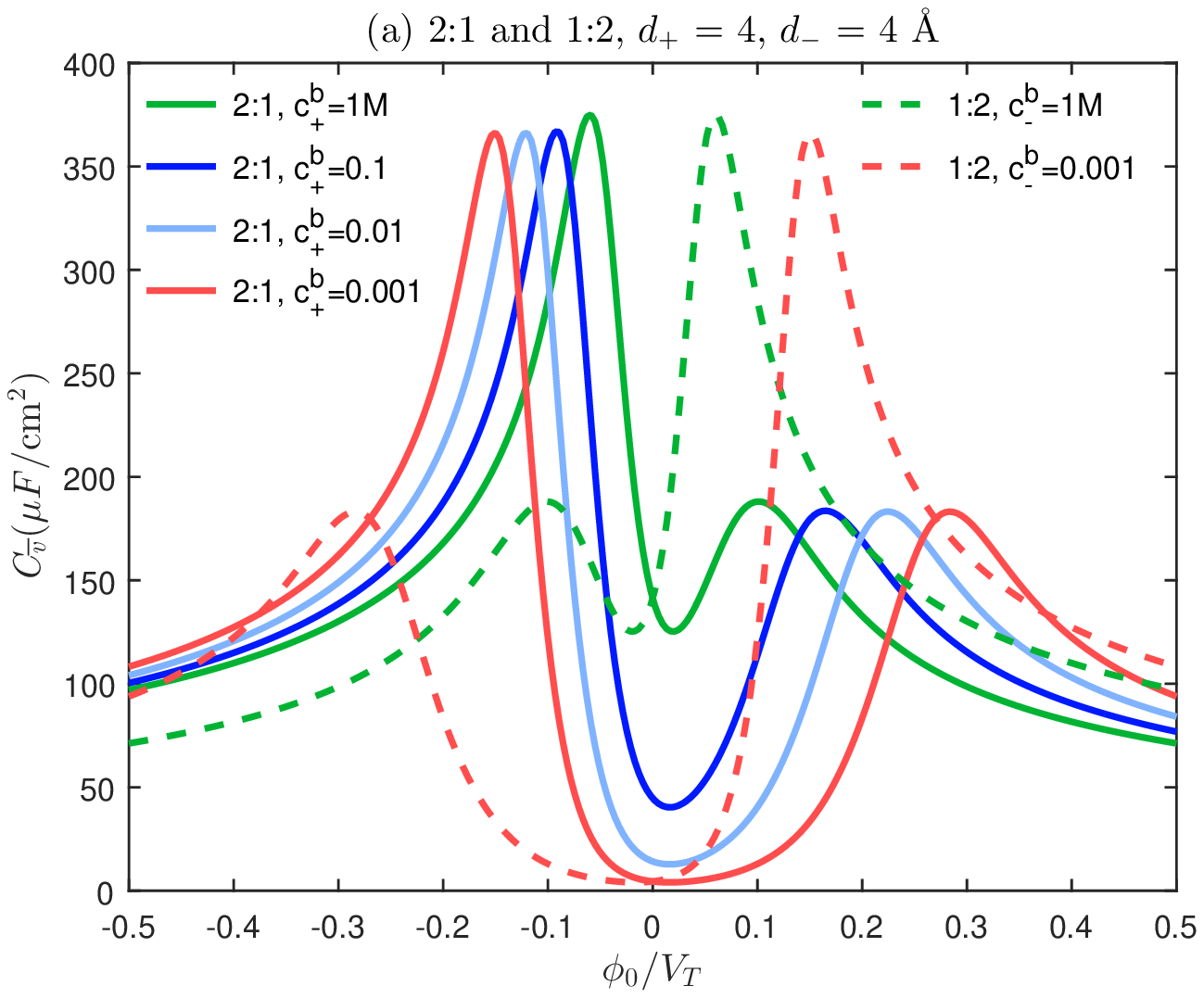}
\end{minipage}\begin{minipage}[t]{0.49\textwidth}
\centering
\includegraphics[width=8cm]{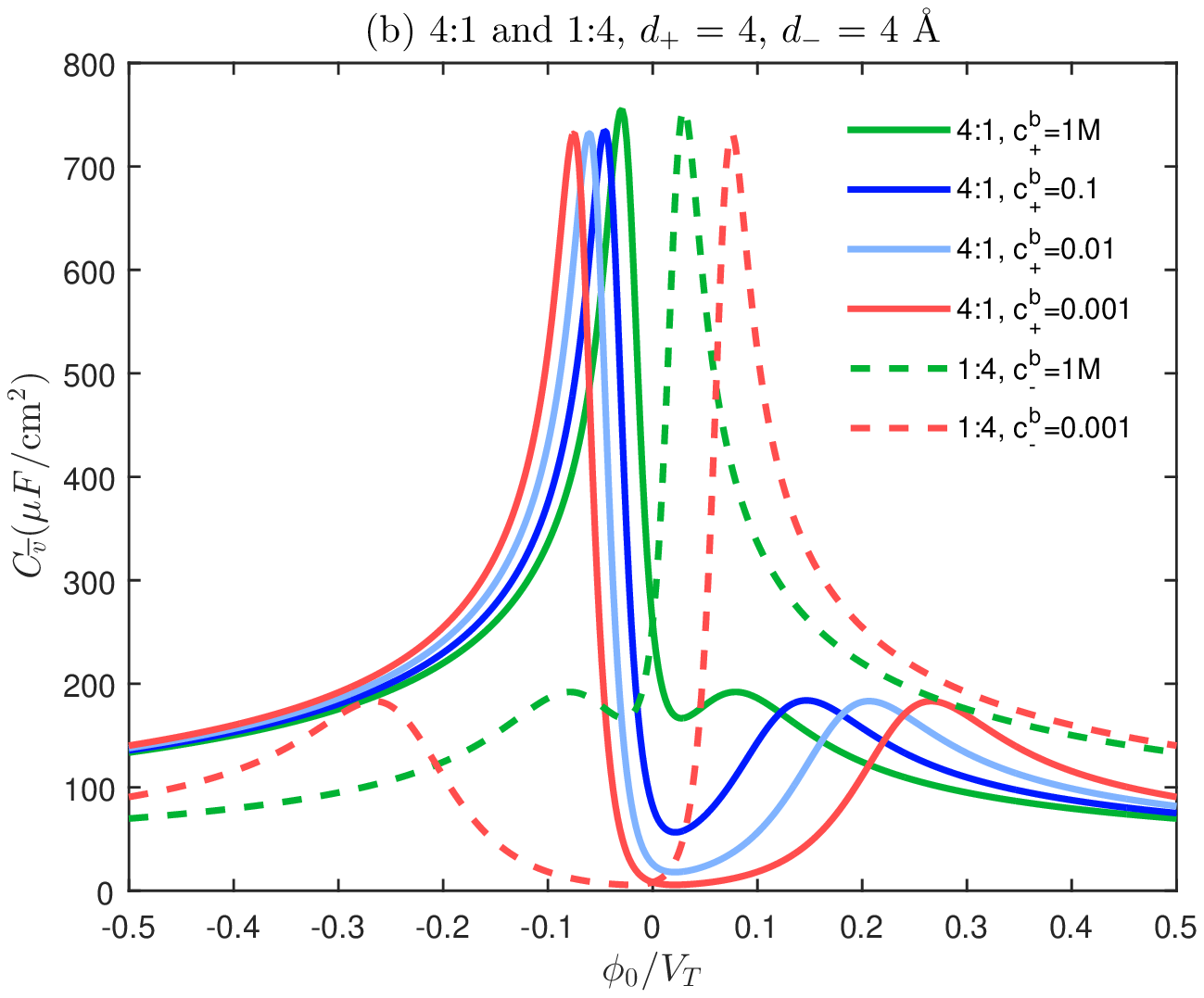}
\end{minipage}
\caption{Asymmetry of DC curves for charge-asymmetric ions in (a) 2:1, 1:2 and
(b) 4:1, 1:4 ionic liquids at various concentrations and fixed $d_{\pm}=4$
{\AA .} }%
\end{figure}

\section{CONCLUSION}

We propose and analyze the differential capacitance (DC) using a formula
derived from the Poisson-Bikerman theory. The formula accounts for varying
steric energies of ions and water in aqueous electrolytes and in ionic liquids
(without water) with different sizes, charges, concentrations, and
compositions. The formula reduces to its classical or contemporary counterpart
when the steric energy vanishes or the ionic size is uniform, displaying the
degenerate nature of models that assume equal diameters, or no volume at all,
or no volume for water molecules. The analysis shows that the differential
capacitance of electrolytes near highly electrified interfaces is determined
by the interplay between electrical and steric energies of ions and water.

The differential capacitance is characterized by critical and inflection
voltages that are defined by a sum of particle volume fractions. It is
important to note that the sum is more general than the mean volume fraction
of equal-sized ions proposed in earlier work. The sum of particle volumes in
the present paper includes \textit{any} arbitrary species of ions and water
with \textit{any} shapes and volumes. The critical and inflection voltages
describe the transition mechanism of DC curves changing from (i) bell shape
(for large values of ionic size or concentration) to (ii) (double-humped
Bactrian) camel shape (for medium value) and then to (iii) U shape (for small
values). The critical voltage also shows that the steric energy of ions and
water becomes more important than the electrical energy of ions in the
adsorbed region of strongly electrified interfaces. The inflection voltage
also describes the saturation property of ions and water in the adsorbed
region as the applied voltage of the electrolyte system reaches to its
physical maximum.

Numerical results illustrate the transition of shapes of DC curves and the
symmetry of experimental DC data for aqueous KPF$_{6}$ electrolyte solution,
which results from delicate interactions of asymmetric ions and water in size.
The asymmetry of DC curves come from the unequal charge of ions of the same
size; or it can come from the unequal size of ions of the same charge. The
more the charge difference is between ions, the larger the difference is
between global maximal and minimal capacitance.

\section{APPENDICES}

\textbf{Appendix A.} We show here the fundamental difference between classical
Poisson-Boltzmann and present Poisson-Bikerman theories that yield Boltzmann
and Fermi distributions of particles, respectively. Substituting eqs 2 and 3
into eq 4 and rearranging terms give
\begin{gather*}
\lbrack c_{i}(\mathbf{r})]^{\overline{v}}=\frac{[c_{i}^{b}\exp(-\beta_{i}%
\phi(\mathbf{r}))]^{\overline{v}}}{(\Gamma^{b})^{v_{i}}}\left(  1-\sum_{j\neq
i}^{K+1}v_{j}c_{j}(\mathbf{r})-v_{i}c_{i}\right)  ^{v_{i}}\\
{}[c_{i}(\mathbf{r})]^{\overline{v}/v_{i}}+\alpha_{i}v_{i}c_{i}(\mathbf{r}%
)=\alpha_{i}\left(  1-\sum_{j\neq i}^{K+1}v_{j}c_{j}(\mathbf{r})\right)  ,
\end{gather*}
where $\alpha_{i}=[c_{i}^{b}\exp(-\beta_{i}\phi(\mathbf{r}))]^{\overline
{v}/v_{i}}/\Gamma^{b}$ for $i=1,\cdots,K+1$. Define $f(t)=t^{\overline
{v}/v_{i}}+\alpha_{i}v_{i}t$ which is an increasing function and satisfies
\[
f(c_{i}(\mathbf{r}))=\alpha_{i}\left(  1-\sum_{j\neq i}^{K+1}v_{j}%
c_{j}(\mathbf{r})\right)  <\alpha_{i}=\alpha_{i}v_{i}\cdot(1/v_{i}%
)<f(1/v_{i}).
\]
Therefore, $c_{i}(\mathbf{r})$ are bounded from above (Fermi distributions),
i.e., $c_{i}(\mathbf{r})$ cannot exceed the maximum value $1/v_{i}$ for any
arbitrary (or even infinite) potential $\phi$ at any location $\mathbf{r}$ in
the domain $\overline{\Omega}$.

In these mean-field Fermi distributions, it is impossible for a volume $v_{i}
$ to be completely filled with ions, i.e., it is impossible to have
$v_{i}c_{i}(\mathbf{r})$ = 1 (and thus $\Gamma(\mathbf{r})=1-v_{i}%
c_{i}(\mathbf{r})=0$ ) since that would make $S(\mathbf{r})=\ln\frac
{\Gamma(\mathbf{r})}{\Gamma^{b}}=-\infty$ and hence $c_{i}(\mathbf{r})=0$, a
contradiction. Therefore, \textit{we must include the void as if it were a
separate species if we treat ions and water having volumes in a model}. This
is a critical property distinguishing our theory from others that do not
consider explicitly the steric potential $S(\mathbf{r})$ with the factor
$v_{i}/\overline{v}$ and the variable volume $v_{i}$ \cite{Liu20}. The steric
potential is consistent with the classical theory of van der Waals in
molecular physics \cite{Liu20}, which describes nonbond interactions between
any pair of atoms as a distance-dependent potential such as the Lennard-Jones
potential that cannot have zero distance between the pair \cite{Liu20}.

In general, we have $\Gamma(\mathbf{r})<\Gamma^{b}$ (or $S(\mathbf{r})<0$)
when $\phi(\mathbf{r})\neq0$ since $c_{i}^{b}\exp(-\beta_{i}\phi
(\mathbf{r}))>c_{i}^{b}$ for $-\beta_{i}\phi(\mathbf{r})>0$ meaning that
negative (positive) $\phi(\mathbf{r})$ at $\mathbf{r}$ attracts positive
(negative) ions of type $i$ to $\mathbf{r}$ and yields $c_{i}(\mathbf{r}%
)>c_{i}^{b}$. We also obtain Boltzmann distributions that diverge if
$-\beta_{i}\phi(\mathbf{r})\rightarrow\infty$ at some $\mathbf{r}$ for
$\beta_{i}\neq0$ (ions not water), i.e., $\lim_{-\beta_{i}\phi(\mathbf{r}%
)\rightarrow\infty}c_{i}^{b}\exp(-\beta_{i}\phi(\mathbf{r})+\frac{v_{i}%
}{\overline{v}}S(\mathbf{r}))=\infty$ when $S(\mathbf{r})=0$, i.e., as
$v_{i}\rightarrow0$ for all $i=1,\cdots,K+1$ (all ions and water are treated
without volumes).

It is important to note that we discovered the importance of the `extra
volume' when we started treating water as a molecule with definite volume. We
could not compute our model without including the extra volume \cite{Liu13a}.
When water was described in a primitive way as a uniform background
dielectric, the computation was not so severely affected. It is possible that
workers using the simple primitive model \cite{Gue10,Val18,Eis10} of classical
fluid theory may have been frustrated by similar unresolved problems.

\textbf{Appendix B.} We now derive eq 7 for equal-sized ions and water, which
is a key formula in the present work and displays a novel and explicit
relationship between electrostatic (the numerator term) and steric
(denominator) potentials. From eqs 5 and 8 with a change of variable
$u=d\phi/dx$, we have
\begin{align}
-\epsilon udu  &  =\rho d\phi=\sum_{i=1}^{K+1}q_{i}c_{i}^{b}\exp(-\beta
_{i}\phi+S)d\phi\nonumber\\
&  =\exp(S)\sum_{i=1}^{K+1}q_{i}c_{i}^{b}\exp(-\beta_{i}\phi)d\phi\tag{B1}\\
&  =\frac{\sum_{i=1}^{K+1}q_{i}c_{i}^{b}\exp(-\beta_{i}\phi)}{\Gamma^{b}%
+\sum_{i=1}^{K+1}v_{i}c_{i}^{b}\exp(-\beta_{i}\phi)}d\phi\tag{B2}%
\end{align}
Introducing $w=\Gamma^{b}+\sum_{i=1}^{K+1}v_{i}c_{i}^{b}\exp(-\beta_{i}\phi)$
and using
\begin{equation}
\frac{dw}{d\phi}=\overline{v}\sum_{i=1}^{K+1}-\beta_{i}c_{i}^{b}\exp
(-\beta_{i}\phi) \tag{B3}%
\end{equation}
yield
\begin{align*}
-\int\epsilon udu  &  =\int\frac{\sum_{i=1}^{K+1}q_{i}c_{i}^{b}\exp(-\beta
_{i}\phi)}{w}d\phi=\int\frac{-k_{B}T/\overline{v}}{w}dw\\
\frac{-\epsilon u^{2}}{2}  &  =\frac{-k_{B}T}{\overline{v}}\ln w+c,
\end{align*}
where $c=0$ since $\phi(\infty)=0$ implying that $w=1$ and $u(\infty)=0$,
i.e., the electric field vanishes at $\infty$. We thus have
\[
u=\frac{d\phi}{dx}=\pm\sqrt{\frac{2k_{B}T}{\epsilon\overline{v}}\ln\left[
\Gamma^{b}+\sum_{i=1}^{K+1}v_{i}c_{i}^{b}\exp(-\beta_{i}\phi)\right]  }.
\]
The boundary condition $-\epsilon\phi^{\prime}(0)=-\epsilon u_{0}=\sigma$
implies that $u_{0}<0$ if $\sigma>0$ and $u_{0}>0$ if $\sigma<0$ and hence
$\pm=\mp sign(\sigma)$. We thus obtain the formula 7 by eq 6, eq 8, and
\begin{align}
C_{\overline{v}}  &  =\mp\epsilon\frac{d}{d\phi_{0}}\left\{  \left(
\frac{2k_{B}T}{\epsilon\overline{v}}\ln\left[  \Gamma^{b}+\sum_{i=1}%
^{K+1}v_{i}c_{i}^{b}\exp(-\beta_{i}\phi_{0})\right]  \right)  ^{1/2}\right\}
\nonumber\\
&  =\pm\epsilon\sqrt{\frac{k_{B}T}{2\epsilon\overline{v}}}\left(  \ln\left[
\Gamma^{b}+\sum_{i=1}^{K+1}v_{i}c_{i}^{b}\exp(-\beta_{i}\phi_{0})\right]
\right)  ^{-1/2}\nonumber\\
&  \frac{\sum_{i=1}^{K+1}\beta_{i}v_{i}c_{i}^{b}\exp(-\beta_{i}\phi_{0}%
)}{\Gamma^{b}+\sum_{i=1}^{K+1}v_{i}c_{i}^{b}\exp(-\beta_{i}\phi_{0})}.
\tag{B4}%
\end{align}
The equal size assumption of particles ($\frac{v_{i}}{\overline{v}}=1$) is a
key to derive eq 7 as shown in eqs B1, B2 by 8, B3, and B4 to 7 by 8.

\textbf{Appendix C.} The differential capacitance eq 10 is for any arbitrary
species of ions and water with any shapes and volumes. The steric term
$\hat{S}$ in eq 10 is a nonlinear function of the surface potential $\phi_{0}$
for which we cannot obtain an explicit formula.

We approximately determine $\hat{S}$ by the following fitting method to the
experimental data $C_{exp}(\phi_{0,k})$ of Valette \cite{Val81}. Setting
$C_{v_{i}}(\phi_{0,k})=C_{exp}(\phi_{0,k})$ yields $\hat{S}_{k}:=\hat{S}%
(\phi_{0,k})$, i.e., $n+1$ values of $\hat{S}$ at $\phi_{0,0}<\phi
_{0,1}<\cdots<\phi_{0,n}=:[a,b]$ for $k=0,\dots,n$. We then use the cubic
spline method to find a function $s(\phi_{0})$ at these $n+1$ interpolation
points such that
\begin{align*}
s_{k}  &  =\hat{S}_{k},k=0,\dots,n\\
s_{k-}^{(j)}  &  =s_{k+}^{(j)},\;k=1,\dots,n-1,\;j=0,1,2,
\end{align*}
where $s_{k}:=s(\phi_{0,k})$ and $s_{k\pm}^{(j)}$ is the $j^{th}$ derivative
of $s(\phi_{0})$ with respect to $\phi_{0}$ at $\phi_{0,k\pm}:=\lim
_{\Delta\phi_{0}\rightarrow0}\phi_{0,k}\pm\Delta\phi_{0}$ . Introducing the
notation $M_{k}:=s_{k}^{(2)}=s_{k\pm}^{(2)}$, the spline function
\[
\begin{aligned} s(\phi_0) &= \frac{ (\phi_{0,k+1} - \phi_0)^3 M_k + (\phi_{0} - \phi_{0,k})^3 M_{k+1} }{6 h_k} + \frac{(\phi_{0,k+1} - \phi_0) \hat{S}_{k} + (\phi_{0} - \phi_{0,k}) \hat{S}_{k+1} }{h_k} \\ &- \frac{h_k}{6} [ (\phi_{0,k+1} - \phi_0) M_k + (\phi_{0} - \phi_{0,k}) M_{k+1} ] \end{aligned}
\]
and its second derivative $s^{(2)}(\phi_{0})$ are continuous functions on the
interval [a, b], where $k=0,1,\dots,n-1,\;\phi_{0}\in\lbrack\phi_{0,k}%
,\phi_{0,k+1}]$, and $h_{k}=\phi_{0,k+1}-\phi_{0,k}$. To determine the unknown
constants $M_{0},\cdots,M_{n}$, we impose $s^{(1)}(\phi_{0})$ to be continuous
in [a, b] as well, i.e.,
\begin{align*}
\frac{h_{k-1}}{6}M_{k-1}+\frac{h_{k}+h_{k-1}}{3}M_{k}+\frac{h_{k}}{6}M_{k+1}
&  =\frac{\hat{S}_{k+1}-\hat{S}_{k}}{h_{k}}-\frac{\hat{S}_{k}-\hat{S}_{k-1}%
}{h_{k-1}}\\
s_{0}^{(1)}  &  =0,\;s_{n}^{(1)}=0.
\end{align*}
for $k=1,\dots,n-1$.

\textbf{Appendix D.} This appendix shows that the DC formula 7 derived from eq
1 is more general than that from the uniform lattice model. The sum of
particle volume fractions $\widetilde{\gamma}=\sum_{i=1}^{K+1}v_{i}%
N_{i}/V=\sum_{i=1}^{K+1}v_{i}c_{i}^{b}$ for arbitrary species of ions and
water with any shapes and volumes is also more general than the mean fraction
formulas of Kornyshev ($\gamma=\overline{N}/N$) \cite{Kor07} and Kilic et al.
($\gamma=2vc^{b}$) \cite{Kil07} with two species ions of the same size.

As a special case, formula 7 reduces to the Bikerman-Freise formula
\cite{Baz09}
\[
C_{BF}=\frac{C_{D}\cosh(\frac{\phi_{0}}{2V_{T}})\sqrt{2\gamma\sinh^{2}%
(\frac{\phi_{0}}{2V_{T}})}}{(1+2\gamma\sinh^{2}\frac{\phi_{0}}{2V_{T}}%
)\sqrt{\ln(1+2\gamma\sinh^{2}\frac{\phi_{0}}{2V_{T}})}}%
\]
also derived by Kornyshev (eq 20 in ref 6) and Kilic et al. \cite{Kil07} if we
replace the mean fraction $\gamma$ with
\begin{equation}
\widetilde{\gamma}=\frac{(v_{1}N_{1}+v_{2}N_{2})}{V}=\frac{v(N_{1}+N_{2}%
)}{(vN)}=\frac{\overline{N}}{N}=2vc^{b}. \tag{D1}%
\end{equation}
Here $\overline{N}=N_{1}+N_{2}$ is the total number of cations $N_{1}$ and
anions $N_{2}$ in the bulk, $N$ is the total number of uniform lattice sites
\cite{Kor07}, $v=8\cdot\frac{3\overline{v}}{4\pi}$ is the volume of each site
(assuming a primitive cubic system), $K=2$, $q_{1}=-q_{2}=e$, $v_{1}%
=v_{2}=v\neq0$, $c^{b}=c_{1}^{b}=c_{2}^{b}$, and $C_{D}=\sqrt{\frac{2\epsilon
e^{2}c^{b}}{k_{B}T}}$ is the Debye capacitance. Note that $\widetilde{\gamma
}=\gamma$ and $\widetilde{\gamma}<\gamma$ for two different volumes $v$ and
$\overline{v}$ of two equal-sized ions occupying their sites without and with
voids, respectively, because $v>\overline{v}$. We assume $v_{1}=v_{2}=v$ in
the general $\widetilde{\gamma}$ as that in the special $\gamma$.

Since probabilities have to sum up to 1, we have
\begin{align*}
1  &  =\Gamma+v(c_{1}+c_{2})\\
&  =\exp(S)\Gamma^{b}+vc^{b}(e^{-\frac{\phi_{0}}{V_{T}}}+e^{\frac{\phi_{0}%
}{V_{T}}})\exp(S)\\
&  =\exp(S)(1-2vc^{b}+2vc^{b}\cosh\frac{\phi_{0}}{V_{T}}).
\end{align*}
This yields the key relation
\begin{equation}
\exp(-S)=1-\gamma+\gamma\cosh(\frac{\phi_{0}}{V_{T}})=1+2\gamma\sinh^{2}%
\frac{\phi_{0}}{2V_{T}}. \tag{D2}%
\end{equation}
Thus,
\begin{align*}
C_{v}  &  =\frac{\pm\overline{C}\sum_{i=1}^{2}q_{i}c_{i}^{b}\exp(-\beta
_{i}\phi_{0})}{\exp(-S_{0})\cdot\sqrt{-S_{0}}}\\
&  =\frac{\pm\sqrt{\frac{\epsilon v}{2k_{B}T}}\cdot ec^{b}(\exp(-\phi
_{0}/V_{T})-\exp(\phi_{0}/V_{T}))}{(1+2\gamma\sinh^{2}\frac{\phi_{0}}{2V_{T}%
})\sqrt{\ln(1+2\gamma\sinh^{2}\frac{\phi_{0}}{2V_{T}})}}\\
&  =\frac{\pm\sqrt{\frac{2\epsilon e^{2}c^{b}}{k_{B}T}}\cdot\frac{\sqrt
{vc^{b}}}{2}(\exp(-\phi_{0}/V_{T})-\exp(\phi_{0}/V_{T}))}{(1+2\gamma\sinh
^{2}\frac{\phi_{0}}{2V_{T}})\sqrt{\ln(1+2\gamma\sinh^{2}\frac{\phi_{0}}%
{2V_{T}})}}\\
&  =\frac{\mp C_{D}\sqrt{\gamma}\sinh(\frac{\phi_{0}}{V_{T}})}{\sqrt
{2}(1+2\gamma\sinh^{2}\frac{\phi_{0}}{2V_{T}})\sqrt{\ln(1+2\gamma\sinh
^{2}\frac{\phi_{0}}{2V_{T}})}},
\end{align*}
and we have the desired result $C_{v}=C_{BF}$ and the Gouy-Chapman law
$\lim_{v\rightarrow0}C_{v}=C_{GC}=C_{D}\cosh(\frac{\phi_{0}}{2V_{T}})$
\cite{Kor07}.

\textbf{Appendix E.} Figure 6 illustrates three curves of the first derivative
$C_{\overline{v}}^{\prime}(\phi_{0})$ of the DC $C_{\overline{v}}(\phi_{0})$
in eq 7 with respect to the surface potential $\phi_{0}$. Those curves yield
critical ($\phi_{0}^{\ast}$) and inflection ($\phi_{0}^{\ast\ast}$) voltages
which describe the transition, space/charge competition, and saturation
properties of ions and water in the condensed layer of EDLs. We explain how to
obtain those curves and voltages.

Differentiating $C_{\overline{v}}(\phi_{0})$ yields the numerator part
\begin{align*}
&  [-\beta_{1}\exp(-\beta_{1}\phi_{0})+\beta_{2}\exp(\beta_{2}\phi_{0}%
)]\exp(-S_{0})\sqrt{-S_{0}}\\
&  -[\exp(-\beta_{1}\phi_{0})-\exp(\beta_{2}\phi_{0})]\left[  \exp
(-S_{0})\sqrt{-S_{0}}+\frac{1}{2}(-S_{0})^{-1/2}\exp(-S_{0})\right]
\frac{d(-S_{0})}{d\phi_{0}}.
\end{align*}
Setting this expression to zero and omitting all the intermediate algebra, we
obtain the equation
\[
-S_{0}\left[  \left(  1-2\overline{v}c^{b}\right)  \left(  t+t^{-1}\right)
+4\overline{v}c^{b}\right]  -\overline{v}c^{b}\left(  t-t^{-1}\right)
^{2}/2=0
\]
for finding critical voltages, i.e., $C_{\overline{v}}^{\prime}(\phi_{0}%
^{\ast})=0$, where $t=\exp(-\beta_{1}\phi_{0})$. If $\phi_{0}=0$, then
$1=\exp(S_{0})\left[  \Gamma^{b}+2\overline{v}c^{b}\right]  =\exp(S_{0})$ by
eqs 2 and 8, $S_{0}=0$, and $C_{\overline{v}}^{\prime}(0)=0$. Therefore, the
DC curve $C_{\overline{v}}(\phi_{0})$ has a local minimum or maximum at
$\phi_{0}^{\ast}=0$. If $\overline{v}=0$ as in the Gouy-Chapman theory, then
$C_{\overline{v}}(\phi_{0})=C_{D}\cosh(\frac{\phi_{0}}{2V_{T}})$ (Appendix D)
is of U shape and has a global minimum at $\phi_{0}^{\ast}=0$. If
$0<\overline{v}c^{b}<1$, \ we can numerically draw the graph of $C_{\overline
{v}}^{\prime}(\phi_{0})$ as shown in Figure 6, then get two inflection
voltages $\pm\phi_{0}^{\ast\ast}$, i.e., $C_{\overline{v}}^{\prime\prime}%
(\pm\phi_{0}^{\ast\ast})=0$, and thus obtain the camel shape of DC curves as
shown in Figures 5a and 5b. If $\overline{v}c^{b}\rightarrow1$, then $\phi
_{0}^{\ast\ast}\rightarrow0=\phi_{0}^{\ast}$ by eq 12, i.e., the inflection
voltage does not exist, $C_{\overline{v}}^{\prime}(\phi_{0})$ has only one
root $\phi_{0}^{\ast}$, and the curve is of bell shape as shown in Figure 5a
for the case of $\overline{v}c^{b}=0.323$ with $\overline{v}=\pi4^{4}/3$
{\AA }$^{3}$ and $c^{b}=2$ M $=2/(1660.6$ {\AA }$^{3})$, for example.

\section{ACKNOWLEDGMENT}

This work was supported by the Ministry of Science and Technology, Taiwan
(MOST 108-2115-M-017-003 to R.-C. C., 109-2115-M-007-011-MY2 to J.-L. L.). We
thank Weishi Liu for valuable comments to improve this work.

\end{document}